\documentclass[preprint]{vgtc}               %

\graphicspath{{figures/}{pictures/}{images/}{./}} %

\usepackage{times}                     %

\usepackage{tasks}

\usepackage{tabu}                      %
\usepackage{booktabs}                  %
\usepackage{lipsum}                    %
\usepackage{mwe}                       %

\usepackage[T1]{fontenc}

\usepackage{mathptmx}                  %

\usepackage{xcolor}
\usepackage{devins_styles}
\usepackage{enumitem}
\setenumerate{noitemsep,topsep=2pt,parsep=0pt,partopsep=2pt}
\setitemize{noitemsep,topsep=2pt,parsep=0pt,partopsep=2pt}
\usepackage{crossreftools}

\onlineid{0}

\vgtccategory{Research}

\vgtcinsertpkg

\title{

A Generative AI System for Biomedical Data Discovery with Grammar-Based Visualizations
}

\author{
Devin Lange\thanks{e-mail: devin@hms.harvard.edu}\\
    \parbox{1.4in}{\scriptsize \centering Harvard Medical School}
\and Shanghua Gao\\
    \scriptsize Harvard Medical School
\and Pengwei Sui\\
    \scriptsize Harvard Medical School
\and Austen Money\\
    \scriptsize Harvard Medical School
\and Priya Misner\\
    \scriptsize Harvard Medical School
\and Marinka Zitnik\\
    \scriptsize Harvard Medical School
\and Nils Gehlenborg\thanks{e-mail: nils@hms.harvard.edu}\\
    \scriptsize Harvard Medical School}

\teaser{
  \centering
  \includegraphics[width=\linewidth]{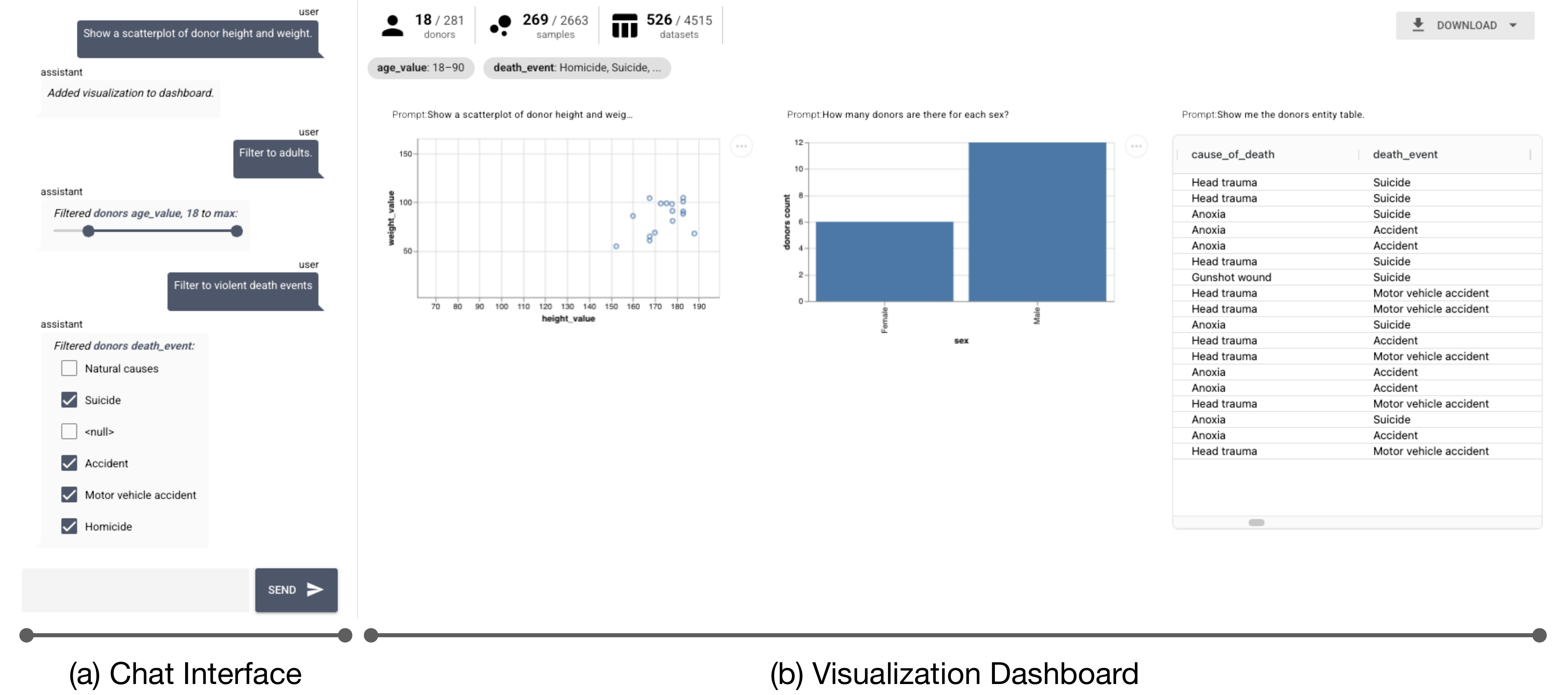}
  \caption{The prototype contains links conversations with progressive visualization dashboards. (a) The chat interface allows users to send free text requests, which result in visualizations and filter actions. Filters also result in UI widgets. (b) The visualization dashboard contains the visualizations, total record counts, current filters, and a download option for the current selection.}
  \label{fig:teaser}
}

\abstract{
We explore the potential for combining generative AI with grammar-based visualizations for biomedical data discovery.
In our prototype, we use a multi-agent system to generate visualization specifications and apply filters. 
These visualizations are linked together, resulting in an interactive dashboard that is progressively constructed.
Our system leverages the strengths of natural language while maintaining the utility of traditional user interfaces. Furthermore, we utilize generated interactive widgets enabling user adjustment.
Finally, we demonstrate the potential utility of this system for biomedical data discovery with a case study.

} %

\begin{document}

\firstsection{Introduction}
\maketitle

Biomedical data discovery is the process by which researchers identify relevant datasets for their scientific inquiry. This step is a prerequisite for discoveries found within biomedical data repositories. Therefore, considerable effort has been devoted to building, deploying, and maintaining sophisticated interfaces for navigating biomedical portals \cite{snyder_human_2019, dekker_4d_2017, roy_elucidating_2023}. However, it is difficult to build an interface that satisfies all scientists' needs. Even after designers and developers have done their due diligence in identifying scientists' needs, they often find additional requests for specific combinations of visualizations or interactions between them.

One approach in developing and maintaining software is to satisfy the most common use cases very well and not worry about the few edge cases that are not possible with the software. This approach aligns well with van Wijk's proposed model for the value of visualization, where more users interacting with visualizations for long periods of time is more valuable \cite{vanwijk_value_2005}.
However, in the setting of biomedical research, a key finding that could lead to a cure for cancer may lie in one of these edge cases.
Finding a way to support the wide diversity of edge cases is critically important and cannot simply be ignored.

This problem touches on a fundamental challenge in designing user interfaces --- there is a tension between the complexity of the interface and its ease of use.
In other words, adding additional features would increase the number of use cases that can technically be accomplished, but also typically increases the difficulty of utilizing those features, sometimes prohibitively so.
What we need is an interface that more flexibly adapts to the users' needs without increasing the burden of interacting with those features.

Natural language interfaces show promise in meeting this need. Users can often express what they want to accomplish in a data discovery environment, even if they are unsure of the correct sequence of actions required within an interface. An interface that can interpret these natural language requests and transform the interface for the user has potential.
With the recent advances in large language models, the visualization community is seeing a renaissance of integration of LLMs with visualization systems \cite{pang_understanding_2025, shen_natural_2023, lyi_learnable_2024, liu_ava_2024, wang_data_2024, lee_sportify_2025, zhu-tian_sporthesia_2023, liu_smartboard_2025, shi_nl2color_2024, yan_knownet_2025, vaithilingam_dynavis_2024, kim_phenoflow_2025, dibia_lida_2023}. These systems have explored various ways in which LLMs can enhance visualization systems. However, none of them have explored the problem of biomedical data discovery.

We developed a prototype application, that integrates a large language model with a visualization interface. However, several things make our interface unique. (1) It interacts with an LLM that has been explicitly fine-tuned for generating interactive biomedical metadata visualizations. (2) It progressively builds multi-view visualizations. (3) These visualizations are automatically linked through a brushing and linking pattern. (4). Our system can generate visualizations across multiple related data tables. (5). Filtering the dataset, which is critical for data discovery, can be accomplished with traditional interactive visualization patterns and from the chatbot. Furthermore, the chatbot also produces UI widgets for the user to adjust or correct the filters that the chatbot created.
Although some of these ideas have been explored in other systems, our prototype is the first interface to integrate all of these for flexible biomedical data discovery.

\section{Related Work}

\subsection{Natural Language Interfaces}

The promise of natural language interfaces is the ability to capitalize on the expressivity and flexibility of natural language to interact with computers. For interfaces that include visualizations, this extends the goal to interacting with data through natural language and visualizations.
Eviza \cite{setlur_eviza_2016} and DataTone \cite{gao_datatone_2015} are two seminal works that set the stage for an outbreak of NLI research \cite{shen_natural_2023} with many interesting directions.
For instance, resolving ambiguity in natural language queries is an important consideration. Both Eviza and DataTone include ambiguity widgets that enable users to adjust the system's initial response. Setlur and Kumar further explored this challenge by using word co-occurrence and sentiment analysis to map vague requests to specific data modifications and presenting those to users in a way that can be modified \cite{setlur_sentifiers_2020}.
More recently, DynaVis allows users to modify visualizations with natural language requests, but also dynamically generates UI widgets for further adjusting the change \cite{vaithilingam_dynavis_2024}.
While DynaVis is designed for visualization authoring, we take a similar approach to resolve ambiguous requests for filtering data during a biomedical data discovery session, thereby resolving ambiguity in users' prompts and rectifying mistakes made by the LLM.

Visualization authoring is a common goal when designing NLI systems \cite{wang_natural_2022, shi_nl2color_2024, lyi_learnable_2024, vaithilingam_dynavis_2024}. There is overlap between data analysis systems and visualization authoring systems; however, our focus is on designing a system to support the specific task of data discovery. The visualizations are a means to an end, not the final product to be created and shared with others.
Alternatively, the general task of interpreting a natural language prompt and generating a visualization is a critical component of our system. Cui et al.\ generate infographics from natural language queries about proportions \cite{cui_text-to-viz_2020}. This work focuses on clear and visually appealing displays of relatively simple information, whereas our application needs to generate many connected interactive visualizations. More similar to our use case is NL4DV, which generates Vega-Lite specifications given a dataset and query \cite{narechania_nl4dv_2021}. %
We use our own fine-tuned LLM to generate visualization specifications for our grammar. Our fine-tuned LLM has been trained to select from multiple data entries or combine tables when generating the visualization specification. Luo et al.\ contribute ncNet, a transformer-based model for generating visualization specifications from natural language \cite {luo_natural_2022}.

In addition to generating initial visualizations with natural language queries, NLIs can be designed to support longer interactive sessions that include follow-up queries or modifications of the current application state. Eviza \cite{setlur_eviza_2016} supported such interactive conversations with data, and Evizeon \cite{hoque_applying_2018} went further by applying language pragmatics to characterize the flow of visual analytical conversations. FlowSense \cite{yu_flowsense_2020} integrated natural language queries into an interface for constructing dataflow diagrams. 
Finally, FlowNL \cite{huang_flownl_2023} enables users to update parameters in a fluid flow visualization using natural language.
While these systems share similarities with our prototype, our primary goal is to facilitate data discovery by progressively building a linked, interactive visualization dashboard that can be filtered using both natural language and traditional user interface interactions.

\subsection{AI for Visualizations}

Artificial intelligence has been used more broadly than just natural language interfaces \cite{wu_ai4vis_2022}.
For instance, Show Me \cite{mackinlay_show_2007} and Voyager \cite{wongsuphasawat_voyager_2016} produce visualization recommendations given an input dataset using rule-based algorithms, and Data2Vis \cite{dibia_data2vis_2018} accomplishes the same thing with a neural network. However, these do not attempt to capture the users' tasks when exploring the data. GenoREC takes this a step further and incorporates user tasks into the genomics visualization recommendation engine \cite{pandey_genorec_2023}. 

Recent advances in large language models have significantly altered the landscape of HCI Research \cite{pang_understanding_2025}. There have been many visualization systems that incorporate large language models in just the last few years \cite{lyi_learnable_2024, liu_ava_2024, wang_data_2024, lee_sportify_2025, zhu-tian_sporthesia_2023, liu_smartboard_2025, shi_nl2color_2024, yan_knownet_2025, vaithilingam_dynavis_2024, kim_phenoflow_2025, dibia_lida_2023, chen_interchat_2025, wang_data_2025, wen_exploring_2025} and research that studies LLMs \cite{wang_dracogpt_2025, cui_promises_2025, chen_viseval_2025}.
The system that shares the most in common with ours is InterChat \cite{chen_interchat_2025}, which allows users to iteratively update a visualization with linked interactions between the visualization and the chat interface. However, there are a few key differences between our prototype and InterChat. InterChat only displays and updates a single visualization, whereas ours displays multiple visualizations simultaneously. Additionally, we support linked filtering of these multiple views through several modalities, interactions with just the visualization, natural language requests, and generated filter widgets. Alternatively, InterChat requires a selection within the visualization and a subsequent natural language request to update the visualization.
For biomedical data discovery, the ability to quickly and precisely filter the data is an essential task and thus prioritized in our design.

Visualizations with LLMs integrated have been developed for domain-specific tasks before. For instance, KNOWNET \cite{yan_knownet_2025} supports health information seeking, Tailor-Mind \cite{gao_fine-tuned_2025} facilitates self-regulated learning, and several tools have been developed for sports analysis \cite{lee_sportify_2025,liu_smartboard_2025, zhu-tian_sporthesia_2023}. However, there have not been any visualization systems that integrate large language models for the task of biomedical data discovery.

Building domain-specific applications sometimes requires fine-tuning large language models \cite{gao2025txagent,gao_fine-tuned_2025}. Fortunately, work has been done to create and publish datasets to facilitate fine-tuning. For instance, nvBench \cite{luo_synthesizing_2021} is a dataset of natural language queries to visualizations. Similarly, Ko et al.\ probide a framework to generate similar datasets \cite{huang_natural-language-based_2020}. We utilize DQVis \cite{lange_dqvis_2025}, a dataset of natural language queries and visualization specifications about biomedical repository metadata to fine-tune our model.

\section{Dataset Description}

We assume that our prototype will operate on a dataset consisting of multiple related entities (tables), and that these tables will contain many quantitative and categorical fields (columns). Furthermore, we assume that these are metadata tables for a data repository, and thus one of the entities will refer to relevant data files, such as genomics data. Hence, the technical goal is to find the list of rows in the dataset entity that can point to additional raw datasets.

\section{Design Requirements}

The primary task we are trying to support with our application is biomedical data discovery. The high-level goal is to find a collection of datasets from within a larger set that meet a set of criteria (both fuzzy and crisp). To guide our development, we created several design requirements.

\begin{enumerate}[align=left]
\setlength\itemsep{0.5em}

\item[{\crtcrossreflabel{\textbf{Filter-Q}}[r:filter-q]}] In some cases, it is necessary to filter the data based on the values of a quantitative field. For instance, only include donors of a specific age. Multiple filters should build on each other. E.g., only donors of a specific age and with a height of over 6 feet.

\item[{\crtcrossreflabel{\textbf{Filter-C}}[r:filter-c]}] Similarly, filtering the data based on categorical values is essential. For instance, only include biological samples that come from a specific organ.

\item[{\crtcrossreflabel{\textbf{Filter-ER}}[r:filter-er]}] Since the goal is to create a list of datasets, the filtering must be applicable across related entities. For instance, filtering donors in a specific age range should also filter biological samples from those donors and datasets derived from those samples.

\item[{\crtcrossreflabel{\textbf{Characterize}}[r:characterize]}] After performing a filtering operation, users must be able to characterize the resulting subset of the data. This step is essential for understanding the result of the filter and for reasoning about whether the resulting subset is acceptable.

\item[{\crtcrossreflabel{\textbf{Transparency}}[r:transparency]}] Since the LLM can perform actions such as filtering, it should be clear when and how the agent does these things. Actions should be transparently communicated in the interface.

\item[{\crtcrossreflabel{\textbf{Resolve}}[r:resolve]}] The user should be able to adjust the actions that the LLM has taken, either to resolve ambiguity in the user's query or to correct a mistake the LLM has made.

\end{enumerate}

\section{Prototype Design}

At its core, our prototype is a linked multi-view visualization that is progressively constructed based on natural language queries. Filters on the data can be progressively explored through interactions with the visualizations and natural language queries. The interface is composed of a conversational chat interface with filter widgets (Figure~
\ref{fig:teaser}a) and a multi-view visualization panel (Figure~\ref{fig:teaser}b).

\begin{figure*}[t]
	\centering
	\includegraphics[width=\linewidth]{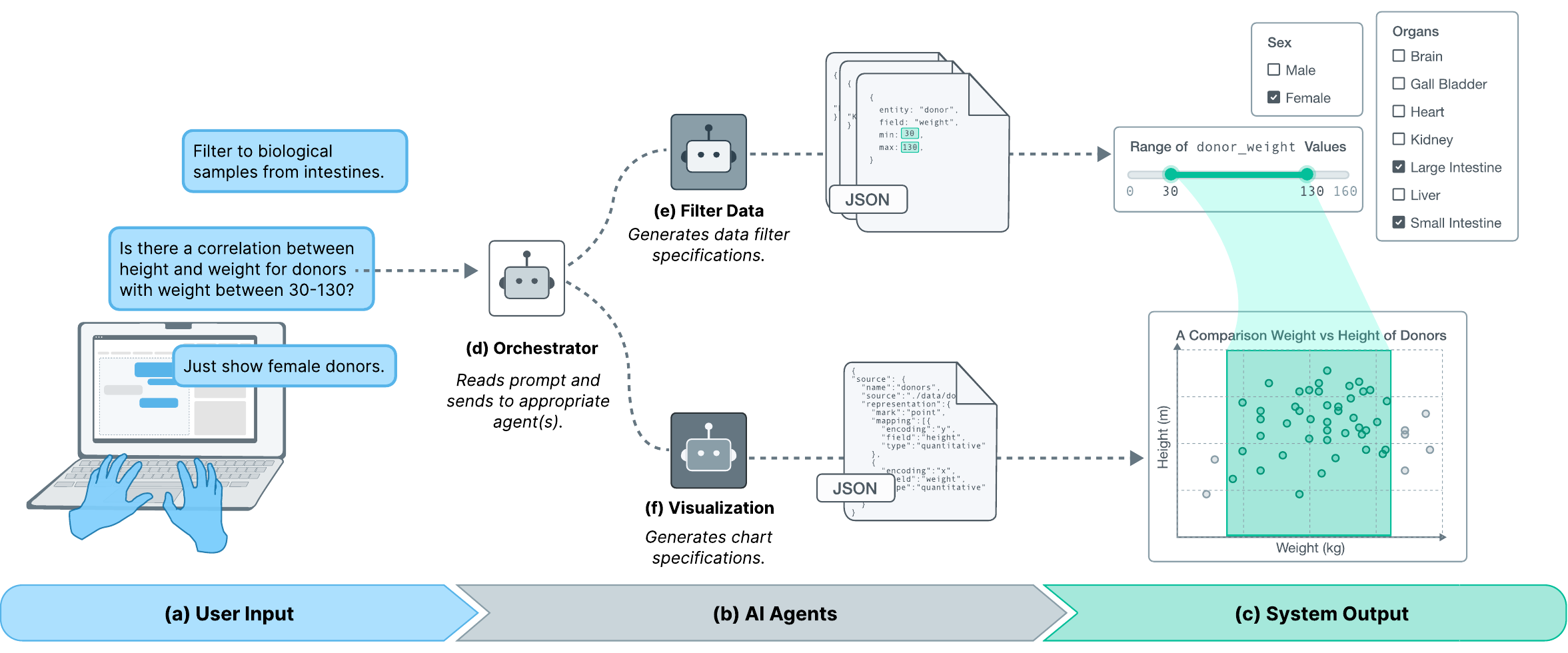}
	\caption{When a user submits a query (a), it is sent to a multi-agent system (b). This system consists of an orchestrator (d), which will then forward the message to a filter agent (d), a visualization agent (f), or both. These will create structured output that will be interpreted and displayed in the system (c).}
	\label{fig:technical_overview}
\end{figure*}

\subsection{Multi-Agent Interaction}
When a user types a message into the chat interface, it gets processed with a multi-agent system composed of three different agents (See Figure~\ref{fig:technical_overview}b). The \textbf{orchestrator} first determines if the user request necessitates filtering the existing data, creating a visualization, or both. Then, additional agents are called. The \textbf{filter} agent will apply new filters to reduce the current dataset, and the \textbf{visualization} agent will produce a new visualization to be added to the dashboard. The delegator and filter agents both make API calls to OpenAI's GPT-4.1 and utilize structured outputs to ensure that the responses are valid. The filter agent can produce interval and point filters. Both types of filters require an entity and a field to operate on, and a range of valid values. \textit{Interval} filters define this range with a minimum and maximum value. Alternatively, \textit{point} filters define a list of valid categorical values. To aid the filter agent, we include the list of entities and fields, and for each field, we include the range of values in the dataset. We exclude categorical fields with many unique values (e.g., ID columns) to avoid exceeding the context window in our request. The visualization agent uses a fine-tuned model. It generates visualization specifications for a grammar that we defined.

\subsection{Visualization Grammar}
We defined a new visualization grammar for biomedical metadata visualizations\footnote{https://hms-dbmi.github.io/udi-grammar}. This grammar takes a grammar of graphics approach to constructing visualizations, and is similar in many ways to Vega-Lite \cite{satyanarayan_vega-lite_2017}. However, there are a few key differences. Our grammar includes additional support for rendering tabular information. It is already possible to render information with Vega-Lite in a grid layout, such as a heatmap; however, displaying tables is not possible. These tabular representations are ubiquitous in data discovery interfaces, so we prioritize them in our grammar. The second distinction is that our grammar supports linking visualizations across multiple specifications, which provides additional flexibility in constructing our multi-view visualizations. We found that using existing models was not very successful at constructing visualization specifications for our grammar. Therefore, we fine-tuned a model to generate these specifications from natural language queries.

\subsection{Fine-Tuning}

To enable the LLM to become an agentic AI model that can generate visualization specifications, we use supervised fine-tuning (SFT) to teach structured tool use. We train on reasoning-and-action traces in which the model decides whether to call a visualization tool by emitting a JSON payload with a function name and arguments, integrates the tool’s output, and then produces the final answer. In SFT, the loss covers both natural-language and tool-call tokens but excludes tool-result tokens to reduce overfitting.
The training dataset comes from human demonstrations and high-quality synthetic rollouts \cite{lange_dqvis_2025}.

\subsection{System Interpretation / Linked interactivity}

The system interprets the structured output of the AI agents. The filter calls result in the data being filtered, and a filter component is displayed in the chat interface and in the dashboard toolbar (\ref{r:transparency}). The filter component is interactive, allowing the user to adjust the initial range selected by the agent (\ref{r:resolve}).

The visualization specifications are rendered as visualization components in the dashboard. Since the fine-tuned LLM was trained to create individual visualizations, the system injects additional logic into the specifications to create interactive linked visualizations. This entails adding the ability to select data within a visualization through 1D intervals, 2D intervals (\ref{r:filter-q}, \ref{r:filter-c}), or point selections, and to display the current data based on the global filter state (\ref{r:characterize}).

Since our system supports multiple entities, we also support linking visualizations across multiple entities (\ref{r:filter-er}). In other words, filtering to donors with an age between 40--60 will also filter to only include biological samples that came from those donors. %

\subsection{Technical Details}
Our prototype is implemented as a web application that utilizes Vue and Quasar for rendering front-end components, as well as our visualization toolkit for creating visualizations. We use VLLM to run the inference of our fine-tuned model, which is exposed through a simple Python API. We run inference on the fine-tuned model using a university HPC, which is suitable for internal development but not for broader deployment.

\section{Case Study}

We illustrate the capabilities of our prototype with a case study of a user data discovery process. First, the user wants to see what metadata is available for the donors table. \userprompt{Show me all the donor data.}. The application produces a basic tabular representation of the donor entity. To more readily understand the distribution of donor metadata, the donor asks \userprompt{How many donors are there for each sex?} and \userprompt{Show a scatterplot of donor height and weight.} This results in two new visualizations being added to the dashboard.
In this case, the user wants donors who are adults, so requests the system \userprompt{filter to adults.} The system then filters the data to donors with an age between 18 and 90. The lower bound of 18 is selected as the most common legal definition of an adult. The upper bound of 90 is chosen because that is the maximum value in this dataset, which was provided as context to the filter agent. However, in this scenario, the user actually wanted donors above the legal drinking age in the US. They use the filter component to adjust the age of donors between 21 and 90 years old. Next, they are interested in patients who died due to violent injuries, so they request \userprompt{filter to violent death events}, resulting in another filter component with several options selected. They also decided to include suicide events in the filter. Finally, they download the identifiers of the donors and the associated datasets.

\section{Discussion}

There are several interesting technical design questions when designing a system like ours. In this section, we describe our current thoughts and challenges while making these decisions.

\subsection{Fine-Tuning versus Multiple Agents}
Our prototype includes calls to both foundational large language models and fine-tuned models. An alternate approach could be to update the fine-tuning to handle both visualizations and filtering. However, it necessitates a longer iteration cycle --- updating the training data and retraining the LLM. This slows the front-end design process, which benefits from faster iteration.
Using a multi-agent system with calls to the foundational model was a faster and more flexible approach while we explored the system design.

\subsection{Evaluation}
Another challenge when fine-tuning an LLM is a robust strategy for evaluating the quality. So far, we have only conducted preliminary ``spot checking'' evaluations of the system. This confirms that the code for fine-tuning ran without errors, and the results are sufficient to develop the initial prototype interface. However, it is not enough to comment on the overall stability or quality of the system.
We plan to conduct a more thorough evaluation of the quality of the fine-tuned LLM, the user interface design, and the entire system.

\subsection{Code Generation Abstraction}
There are several possible strategies for LLMs to generate visualizations at different levels of abstraction.
For instance, it can generate code (e.g., Python or JavaScript) that must be executed to create the visualization. Alternatively, it can generate a specification (e.g., JSON) that provides all the necessary information to create the visualization. For our project, we decided to target specifications in the form of a grammar we designed. We believe there are several advantages to this choice.
The first advantage is the guardrails that this approach provides. The visualization agent is designed to create only visualization specifications.
The interface will ignore anything generated outside the bounds of our JSON Schema.
The second advantage is that the structured nature of the specification outputs facilitates the integration of responses into a larger system. For instance, our fine-tuned model was trained to generate a single visualization in response to a prompt.
However, due to the well-defined structure, the system can inject interactivity into these specifications, resulting in a dashboard of linked visualizations.

\section{Conclusion and Future Work}

In conclusion, we developed a prototype that integrates elements of a traditional data discovery interface with a chat-based system. It utilizes a fine-tuned LLM to generate visualizations that progressively add to a multi-view visualization. These visualizations are automatically interactive with a linked brushing and filtering pattern. The system supports data schemas for multiple related entities, and interactions within the system are facilitated with parity, transparency, and resolution of LLM actions through filter widgets.

The next critical piece of work is a more careful evaluation of the fine-tuned model, the interface design, and the integrated system. With a more robust evaluation in place, we will continue to refine the interface, fine-tuned model, and multi-agent design for interacting with the system.

\acknowledgments{
This work was supported by ARPA-H AY2AX000028. The authors wish to thank members of the HIDIVE lab for their support and feedback throughout this project. 
}

\bibliographystyle{abbrv-doi}

\bibliography{ref, manual_refs}

\begin{thebibliography}{10}

\bibitem{chen_interchat_2025}
J.~Chen, J.~Wu, J.~Guo, V.~Mohanty, X.~Li, J.~P. Ono, W.~He, L.~Ren, and D.~Liu.
\newblock {{InterChat}}: {{Enhancing Generative Visual Analytics}} using {{Multimodal Interactions}}.
\newblock {\em Computer Graphics Forum}, 44(3):e70112, 2025. doi: {{%
10\hspace{.1pt}\discretionary{.}{%
}{.}\hspace{.4pt}1111\discretionary{/}{%
}{/}cgf\hspace{.1pt}\discretionary{.}{%
}{.}\hspace{.4pt}70112}}


\bibitem{chen_viseval_2025}
N.~Chen, Y.~Zhang, J.~Xu, K.~Ren, and Y.~Yang.
\newblock {{VisEval}}: {{A Benchmark}} for {{Data Visualization}} in the {{Era}} of {{Large Language Models}}.
\newblock {\em IEEE Transactions on Visualization and Computer Graphics}, 31(1):1301--1311, Jan. 2025. doi: {{%
10\hspace{.1pt}\discretionary{.}{%
}{.}\hspace{.4pt}1109\discretionary{/}{%
}{/}TVCG\hspace{.1pt}\discretionary{.}{%
}{.}\hspace{.4pt}2024\hspace{.1pt}\discretionary{.}{%
}{.}\hspace{.4pt}3456320}}


\bibitem{cui_text-to-viz_2020}
W.~Cui, X.~Zhang, Y.~Wang, H.~Huang, B.~Chen, L.~Fang, H.~Zhang, J.-G. Lou, and D.~Zhang.
\newblock Text-to-{{Viz}}: {{Automatic Generation}} of {{Infographics}} from {{Proportion-Related Natural Language Statements}}.
\newblock {\em IEEE Transactions on Visualization and Computer Graphics}, 26(1):906--916, Jan. 2020. doi: {{%
10\hspace{.1pt}\discretionary{.}{%
}{.}\hspace{.4pt}1109\discretionary{/}{%
}{/}TVCG\hspace{.1pt}\discretionary{.}{%
}{.}\hspace{.4pt}2019\hspace{.1pt}\discretionary{.}{%
}{.}\hspace{.4pt}2934785}}


\bibitem{cui_promises_2025}
Y.~Cui, L.~W. Ge, Y.~Ding, L.~Harrison, F.~Yang, and M.~Kay.
\newblock Promises and {{Pitfalls}}: {{Using Large Language Models}} to {{Generate Visualization Items}}.
\newblock {\em IEEE Transactions on Visualization and Computer Graphics}, 31(1):1094--1104, Jan. 2025. doi: {{%
10\hspace{.1pt}\discretionary{.}{%
}{.}\hspace{.4pt}1109\discretionary{/}{%
}{/}TVCG\hspace{.1pt}\discretionary{.}{%
}{.}\hspace{.4pt}2024\hspace{.1pt}\discretionary{.}{%
}{.}\hspace{.4pt}3456309}}


\bibitem{dekker_4d_2017}
J.~Dekker, A.~S. Belmont, M.~Guttman, V.~O. Leshyk, J.~T. Lis, S.~Lomvardas, L.~A. Mirny, C.~C. O'Shea, P.~J. Park, B.~Ren, J.~C.~R. Politz, J.~Shendure, and S.~Zhong.
\newblock The {{4D}} nucleome project.
\newblock {\em Nature}, 549(7671):219--226, Sept. 2017. doi: {{%
10\hspace{.1pt}\discretionary{.}{%
}{.}\hspace{.4pt}1038\discretionary{/}{%
}{/}nature23884}}


\bibitem{dibia_lida_2023}
V.~Dibia.
\newblock {{LIDA}}: {{A Tool}} for {{Automatic Generation}} of {{Grammar-Agnostic Visualizations}} and {{Infographics}} using {{Large Language Models}}.
\newblock In D.~Bollegala, R.~Huang, and A.~Ritter, eds., {\em Proceedings of the 61st {{Annual Meeting}} of the {{Association}} for {{Computational Linguistics}} ({{Volume}} 3: {{System Demonstrations}})}, pp. 113--126. Association for Computational Linguistics, Toronto, Canada, July 2023. doi: {{%
10\hspace{.1pt}\discretionary{.}{%
}{.}\hspace{.4pt}18653\discretionary{/}{%
}{/}v1\discretionary{/}{%
}{/}2023\hspace{.1pt}\discretionary{.}{%
}{.}\hspace{.4pt}acl\discretionary{%
}{-}{-}demo\hspace{.1pt}\discretionary{.}{%
}{.}\hspace{.4pt}11}}


\bibitem{dibia_data2vis_2018}
V.~Dibia and {\c C}.~Demiralp.
\newblock {{Data2Vis}}: {{Automatic Generation}} of {{Data Visualizations Using Sequence}} to {{Sequence Recurrent Neural Networks}}, Nov. 2018. doi: {{%
10\hspace{.1pt}\discretionary{.}{%
}{.}\hspace{.4pt}48550\discretionary{/}{%
}{/}arXiv\hspace{.1pt}\discretionary{.}{%
}{.}\hspace{.4pt}1804\hspace{.1pt}\discretionary{.}{%
}{.}\hspace{.4pt}03126}}


\bibitem{gao_fine-tuned_2025}
L.~Gao, J.~Lu, Z.~Shao, Z.~Lin, S.~Yue, C.~Leong, Y.~Sun, R.~J. Zauner, Z.~Wei, and S.~Chen.
\newblock Fine-{{Tuned Large Language Model}} for {{Visualization System}}: {{A Study}} on {{Self-Regulated Learning}} in {{Education}}.
\newblock {\em IEEE Transactions on Visualization and Computer Graphics}, 31(1):514--524, Jan. 2025. doi: {{%
10\hspace{.1pt}\discretionary{.}{%
}{.}\hspace{.4pt}1109\discretionary{/}{%
}{/}TVCG\hspace{.1pt}\discretionary{.}{%
}{.}\hspace{.4pt}2024\hspace{.1pt}\discretionary{.}{%
}{.}\hspace{.4pt}3456145}}


\bibitem{gao2025txagent}
S.~Gao, R.~Zhu, Z.~Kong, A.~Noori, X.~Su, C.~Ginder, T.~Tsiligkaridis, and M.~Zitnik.
\newblock Txagent: An ai agent for therapeutic reasoning across a universe of tools, 2025.

\bibitem{gao_datatone_2015}
T.~Gao, M.~Dontcheva, E.~Adar, Z.~Liu, and K.~G. Karahalios.
\newblock {{DataTone}}: {{Managing Ambiguity}} in {{Natural Language Interfaces}} for {{Data Visualization}}.
\newblock In {\em Proceedings of the 28th {{Annual ACM Symposium}} on {{User Interface Software}} \& {{Technology}}}, pp. 489--500. ACM, Charlotte NC USA, Nov. 2015. doi: {{%
10\hspace{.1pt}\discretionary{.}{%
}{.}\hspace{.4pt}1145\discretionary{/}{%
}{/}2807442\hspace{.1pt}\discretionary{.}{%
}{.}\hspace{.4pt}2807478}}


\bibitem{hoque_applying_2018}
E.~Hoque, V.~Setlur, M.~Tory, and I.~Dykeman.
\newblock Applying {{Pragmatics Principles}} for {{Interaction}} with {{Visual Analytics}}.
\newblock {\em IEEE Transactions on Visualization and Computer Graphics}, 24(1):309--318, Jan. 2018. doi: {{%
10\hspace{.1pt}\discretionary{.}{%
}{.}\hspace{.4pt}1109\discretionary{/}{%
}{/}TVCG\hspace{.1pt}\discretionary{.}{%
}{.}\hspace{.4pt}2017\hspace{.1pt}\discretionary{.}{%
}{.}\hspace{.4pt}2744684}}


\bibitem{huang_flownl_2023}
J.~Huang, Y.~Xi, J.~Hu, and J.~Tao.
\newblock {{FlowNL}}: {{Asking}} the {{Flow Data}} in {{Natural Languages}}.
\newblock {\em IEEE Transactions on Visualization and Computer Graphics}, 29(1):1200--1210, Jan. 2023. doi: {{%
10\hspace{.1pt}\discretionary{.}{%
}{.}\hspace{.4pt}1109\discretionary{/}{%
}{/}TVCG\hspace{.1pt}\discretionary{.}{%
}{.}\hspace{.4pt}2022\hspace{.1pt}\discretionary{.}{%
}{.}\hspace{.4pt}3209453}}


\bibitem{huang_natural-language-based_2020}
Z.~Huang, Y.~Zhao, W.~Chen, S.~Gao, K.~Yu, W.~Xu, M.~Tang, M.~Zhu, and M.~Xu.
\newblock A {{Natural-language-based Visual Query Approach}} of {{Uncertain Human Trajectories}}.
\newblock {\em IEEE Transactions on Visualization and Computer Graphics}, 26(1):1256--1266, Jan. 2020. doi: {{%
10\hspace{.1pt}\discretionary{.}{%
}{.}\hspace{.4pt}1109\discretionary{/}{%
}{/}TVCG\hspace{.1pt}\discretionary{.}{%
}{.}\hspace{.4pt}2019\hspace{.1pt}\discretionary{.}{%
}{.}\hspace{.4pt}2934671}}


\bibitem{kim_phenoflow_2025}
J.~Kim, S.~Lee, H.~Jeon, K.-J. Lee, H.-J. Bae, B.~Kim, and J.~Seo.
\newblock {{PhenoFlow}}: {{A Human-LLM Driven Visual Analytics System}} for {{Exploring Large}} and {{Complex Stroke Datasets}}.
\newblock {\em IEEE Transactions on Visualization and Computer Graphics}, 31(1):470--480, Jan. 2025. doi: {{%
10\hspace{.1pt}\discretionary{.}{%
}{.}\hspace{.4pt}1109\discretionary{/}{%
}{/}TVCG\hspace{.1pt}\discretionary{.}{%
}{.}\hspace{.4pt}2024\hspace{.1pt}\discretionary{.}{%
}{.}\hspace{.4pt}3456215}}


\bibitem{lange_dqvis_2025}
D.~Lange, P.~Sui, S.~Gao, M.~Zitnik, and N.~Gehlenborg.
\newblock {{DQVis Dataset}}: {{Natural Language}} to {{Biomedical Visualization}}, 2025.

\bibitem{lee_sportify_2025}
C.~Lee, T.~Lin, H.~Pfister, and C.~{Zhu-Tian}.
\newblock Sportify: {{Question Answering}} with {{Embedded Visualizations}} and {{Personified Narratives}} for {{Sports Video}}.
\newblock {\em IEEE Transactions on Visualization and Computer Graphics}, 31(1):12--22, Jan. 2025. doi: {{%
10\hspace{.1pt}\discretionary{.}{%
}{.}\hspace{.4pt}1109\discretionary{/}{%
}{/}TVCG\hspace{.1pt}\discretionary{.}{%
}{.}\hspace{.4pt}2024\hspace{.1pt}\discretionary{.}{%
}{.}\hspace{.4pt}3456332}}


\bibitem{liu_ava_2024}
S.~Liu, H.~Miao, Z.~Li, M.~Olson, V.~Pascucci, and P.-T. Bremer.
\newblock {{AVA}}: {{Towards Autonomous Visualization Agents}} through {{Visual Perception-Driven Decision-Making}}.
\newblock {\em Computer Graphics Forum}, 43(3):e15093, 2024. doi: {{%
10\hspace{.1pt}\discretionary{.}{%
}{.}\hspace{.4pt}1111\discretionary{/}{%
}{/}cgf\hspace{.1pt}\discretionary{.}{%
}{.}\hspace{.4pt}15093}}


\bibitem{liu_smartboard_2025}
Z.~Liu, X.~Xie, M.~He, W.~Zhao, Y.~Wu, L.~Cheng, H.~Zhang, and Y.~Wu.
\newblock Smartboard: {{Visual Exploration}} of {{Team Tactics}} with {{LLM Agent}}.
\newblock {\em IEEE Transactions on Visualization and Computer Graphics}, 31(1):23--33, Jan. 2025. doi: {{%
10\hspace{.1pt}\discretionary{.}{%
}{.}\hspace{.4pt}1109\discretionary{/}{%
}{/}TVCG\hspace{.1pt}\discretionary{.}{%
}{.}\hspace{.4pt}2024\hspace{.1pt}\discretionary{.}{%
}{.}\hspace{.4pt}3456200}}


\bibitem{luo_synthesizing_2021}
Y.~Luo, N.~Tang, G.~Li, C.~Chai, W.~Li, and X.~Qin.
\newblock Synthesizing {{Natural Language}} to {{Visualization}} ({{NL2VIS}}) {{Benchmarks}} from {{NL2SQL Benchmarks}}.
\newblock In {\em Proceedings of the 2021 {{International Conference}} on {{Management}} of {{Data}}}, {{SIGMOD}} '21, pp. 1235--1247. Association for Computing Machinery, New York, NY, USA, June 2021. doi: {{%
10\hspace{.1pt}\discretionary{.}{%
}{.}\hspace{.4pt}1145\discretionary{/}{%
}{/}3448016\hspace{.1pt}\discretionary{.}{%
}{.}\hspace{.4pt}3457261}}


\bibitem{luo_natural_2022}
Y.~Luo, N.~Tang, G.~Li, J.~Tang, C.~Chai, and X.~Qin.
\newblock Natural {{Language}} to {{Visualization}} by {{Neural Machine Translation}}.
\newblock {\em IEEE Transactions on Visualization and Computer Graphics}, 28(1):217--226, Jan. 2022. doi: {{%
10\hspace{.1pt}\discretionary{.}{%
}{.}\hspace{.4pt}1109\discretionary{/}{%
}{/}TVCG\hspace{.1pt}\discretionary{.}{%
}{.}\hspace{.4pt}2021\hspace{.1pt}\discretionary{.}{%
}{.}\hspace{.4pt}3114848}}


\bibitem{lyi_learnable_2024}
S.~L'Yi, A.~{van den Brandt}, E.~Adams, H.~N. Nguyen, and N.~Gehlenborg.
\newblock Learnable and {{Expressive Visualization Authoring}} through {{Blended Interfaces}}.
\newblock {\em IEEE Transactions on Visualization and Computer Graphics}, 2024. doi: {{%
10\hspace{.1pt}\discretionary{.}{%
}{.}\hspace{.4pt}1109\discretionary{/}{%
}{/}TVCG\hspace{.1pt}\discretionary{.}{%
}{.}\hspace{.4pt}2024\hspace{.1pt}\discretionary{.}{%
}{.}\hspace{.4pt}3456598}}


\bibitem{mackinlay_show_2007}
J.~Mackinlay, P.~Hanrahan, and C.~Stolte.
\newblock Show {{Me}}: {{Automatic Presentation}} for {{Visual Analysis}}.
\newblock {\em IEEE Transactions on Visualization and Computer Graphics (InfoVis '07)}, 13(6):1137--1144, 2007. doi: {{%
10\hspace{.1pt}\discretionary{.}{%
}{.}\hspace{.4pt}1109\discretionary{/}{%
}{/}TVCG\hspace{.1pt}\discretionary{.}{%
}{.}\hspace{.4pt}2007\hspace{.1pt}\discretionary{.}{%
}{.}\hspace{.4pt}70594}}


\bibitem{narechania_nl4dv_2021}
A.~Narechania, A.~Srinivasan, and J.~Stasko.
\newblock {{NL4DV}}: {{A Toolkit}} for {{Generating Analytic Specifications}} for {{Data Visualization}} from {{Natural Language Queries}}.
\newblock {\em IEEE Transactions on Visualization and Computer Graphics}, 27(2):369--379, Feb. 2021. doi: {{%
10\hspace{.1pt}\discretionary{.}{%
}{.}\hspace{.4pt}1109\discretionary{/}{%
}{/}TVCG\hspace{.1pt}\discretionary{.}{%
}{.}\hspace{.4pt}2020\hspace{.1pt}\discretionary{.}{%
}{.}\hspace{.4pt}3030378}}


\bibitem{pandey_genorec_2023}
A.~Pandey, S.~L'Yi, Q.~Wang, M.~A. Borkin, and N.~Gehlenborg.
\newblock {{GenoREC}}: {{A Recommendation System}} for {{Interactive Genomics Data Visualization}}.
\newblock {\em IEEE Transactions on Visualization and Computer Graphics}, 29(1):570--580, Jan. 2023. doi: {{%
10\hspace{.1pt}\discretionary{.}{%
}{.}\hspace{.4pt}1109\discretionary{/}{%
}{/}TVCG\hspace{.1pt}\discretionary{.}{%
}{.}\hspace{.4pt}2022\hspace{.1pt}\discretionary{.}{%
}{.}\hspace{.4pt}3209407}}


\bibitem{pang_understanding_2025}
R.~Y. Pang, H.~Schroeder, K.~S. Smith, S.~Barocas, Z.~Xiao, E.~Tseng, and D.~Bragg.
\newblock Understanding the {{LLM-ification}} of {{CHI}}: {{Unpacking}} the {{Impact}} of {{LLMs}} at {{CHI}} through a {{Systematic Literature Review}}.
\newblock In {\em Proceedings of the 2025 {{CHI Conference}} on {{Human Factors}} in {{Computing Systems}}}, {{CHI}} '25, pp. 1--20. Association for Computing Machinery, New York, NY, USA, Apr. 2025. doi: {{%
10\hspace{.1pt}\discretionary{.}{%
}{.}\hspace{.4pt}1145\discretionary{/}{%
}{/}3706598\hspace{.1pt}\discretionary{.}{%
}{.}\hspace{.4pt}3713726}}


\bibitem{roy_elucidating_2023}
A.~L. Roy, R.~S. Conroy, V.~G. Taylor, J.~Mietz, I.~M. Fingerman, M.~J. Pazin, P.~Smith, C.~M. Hutter, D.~S. Singer, and E.~L. Wilder.
\newblock Elucidating the structure and function of the nucleus---{{The NIH Common Fund 4D Nucleome}} program.
\newblock {\em Molecular Cell}, 83(3):335--342, Feb. 2023. doi: {{%
10\hspace{.1pt}\discretionary{.}{%
}{.}\hspace{.4pt}1016\discretionary{/}{%
}{/}j\hspace{.1pt}\discretionary{.}{%
}{.}\hspace{.4pt}molcel\hspace{.1pt}\discretionary{.}{%
}{.}\hspace{.4pt}2022\hspace{.1pt}\discretionary{.}{%
}{.}\hspace{.4pt}12\hspace{.1pt}\discretionary{.}{%
}{.}\hspace{.4pt}025}}


\bibitem{satyanarayan_vega-lite_2017}
A.~Satyanarayan, D.~Moritz, K.~Wongsuphasawat, and J.~Heer.
\newblock Vega-{{Lite}}: {{A Grammar}} of {{Interactive Graphics}}.
\newblock {\em IEEE Transactions on Visualization and Computer Graphics}, 23(1):341--350, Jan. 2017. doi: {{%
10\hspace{.1pt}\discretionary{.}{%
}{.}\hspace{.4pt}1109\discretionary{/}{%
}{/}TVCG\hspace{.1pt}\discretionary{.}{%
}{.}\hspace{.4pt}2016\hspace{.1pt}\discretionary{.}{%
}{.}\hspace{.4pt}2599030}}


\bibitem{setlur_eviza_2016}
V.~Setlur, S.~E. Battersby, M.~Tory, R.~Gossweiler, and A.~X. Chang.
\newblock Eviza: {{A Natural Language Interface}} for {{Visual Analysis}}.
\newblock In {\em Proceedings of the 29th {{Annual Symposium}} on {{User Interface Software}} and {{Technology}}}, {{UIST}} '16, pp. 365--377. Association for Computing Machinery, New York, NY, USA, Oct. 2016. doi: {{%
10\hspace{.1pt}\discretionary{.}{%
}{.}\hspace{.4pt}1145\discretionary{/}{%
}{/}2984511\hspace{.1pt}\discretionary{.}{%
}{.}\hspace{.4pt}2984588}}


\bibitem{setlur_sentifiers_2020}
V.~Setlur and A.~Kumar.
\newblock Sentifiers: {{Interpreting Vague Intent Modifiers}} in {{Visual Analysis}} using {{Word Co-occurrence}} and {{Sentiment Analysis}}.
\newblock In {\em 2020 {{IEEE Visualization Conference}} ({{VIS}})}, pp. 216--220, Oct. 2020. doi: {{%
10\hspace{.1pt}\discretionary{.}{%
}{.}\hspace{.4pt}1109\discretionary{/}{%
}{/}VIS47514\hspace{.1pt}\discretionary{.}{%
}{.}\hspace{.4pt}2020\hspace{.1pt}\discretionary{.}{%
}{.}\hspace{.4pt}00050}}


\bibitem{shen_natural_2023}
L.~Shen, E.~Shen, Y.~Luo, X.~Yang, X.~Hu, X.~Zhang, Z.~Tai, and J.~Wang.
\newblock Towards {{Natural Language Interfaces}} for {{Data Visualization}}: {{A Survey}}.
\newblock {\em IEEE Transactions on Visualization and Computer Graphics}, 29(6):3121--3144, June 2023. doi: {{%
10\hspace{.1pt}\discretionary{.}{%
}{.}\hspace{.4pt}1109\discretionary{/}{%
}{/}TVCG\hspace{.1pt}\discretionary{.}{%
}{.}\hspace{.4pt}2022\hspace{.1pt}\discretionary{.}{%
}{.}\hspace{.4pt}3148007}}


\bibitem{shi_nl2color_2024}
C.~Shi, W.~Cui, C.~Liu, C.~Zheng, H.~Zhang, Q.~Luo, and X.~Ma.
\newblock {{NL2Color}}: {{Refining Color Palettes}} for {{Charts}} with {{Natural Language}}.
\newblock {\em IEEE Transactions on Visualization and Computer Graphics}, 30(1):814--824, Jan. 2024. doi: {{%
10\hspace{.1pt}\discretionary{.}{%
}{.}\hspace{.4pt}1109\discretionary{/}{%
}{/}TVCG\hspace{.1pt}\discretionary{.}{%
}{.}\hspace{.4pt}2023\hspace{.1pt}\discretionary{.}{%
}{.}\hspace{.4pt}3326522}}


\bibitem{snyder_human_2019}
M.~P. Snyder, S.~Lin, A.~Posgai, M.~Atkinson, A.~Regev, J.~Rood, O.~{Rozenblatt-Rosen}, L.~Gaffney, A.~Hupalowska, R.~Satija, N.~Gehlenborg, J.~Shendure, J.~Laskin, P.~Harbury, N.~A. Nystrom, J.~C. Silverstein, Z.~{Bar-Joseph}, K.~Zhang, K.~B{\"o}rner, Y.~Lin, R.~Conroy, D.~Procaccini, A.~L. Roy, A.~Pillai, M.~Brown, Z.~S. Galis, L.~Cai, J.~Shendure, C.~Trapnell, S.~Lin, D.~Jackson, M.~P. Snyder, G.~Nolan, W.~J. Greenleaf, Y.~Lin, S.~Plevritis, S.~Ahadi, S.~A. Nevins, H.~Lee, C.~M. Schuerch, S.~Black, V.~G. Venkataraaman, E.~Esplin, A.~Horning, A.~Bahmani, K.~Zhang, X.~Sun, S.~Jain, J.~Hagood, G.~Pryhuber, P.~Kharchenko, M.~Atkinson, B.~Bodenmiller, T.~Brusko, M.~{Clare-Salzler}, H.~Nick, K.~Otto, A.~Posgai, C.~Wasserfall, M.~Jorgensen, M.~Brusko, S.~Maffioletti, R.~M. Caprioli, J.~M. Spraggins, D.~Gutierrez, N.~H. Patterson, E.~K. Neumann, R.~Harris, M.~{deCaestecker}, A.~B. Fogo, R.~{van de Plas}, K.~Lau, L.~Cai, G.-C. Yuan, Q.~Zhu, R.~Dries, P.~Yin, S.~K. Saka, J.~Y. Kishi, Y.~Wang, I.~Goldaracena,
  J.~Laskin, D.~Ye, K.~E. {Burnum-Johnson}, P.~D. Piehowski, C.~Ansong, Y.~Zhu, P.~Harbury, T.~Desai, J.~Mulye, P.~Chou, M.~Nagendran, Z.~{Bar-Joseph}, S.~A. Teichmann, B.~Paten, R.~F. Murphy, J.~Ma, V.~Y. Kiselev, C.~Kingsford, A.~Ricarte, M.~Keays, S.~A. Akoju, M.~Ruffalo, N.~Gehlenborg, P.~Kharchenko, M.~Vella, C.~McCallum, K.~B{\"o}rner, L.~E. Cross, S.~H. Friedman, R.~Heiland, B.~Herr, P.~Macklin, E.~M. Quardokus, L.~Record, J.~P. Sluka, G.~M. Weber, N.~A. Nystrom, J.~C. Silverstein, P.~D. Blood, A.~J. Ropelewski, W.~E. Shirey, R.~M. Scibek, P.~Mabee, W.~C. Lenhardt, K.~Robasky, S.~Michailidis, R.~Satija, J.~Marioni, A.~Regev, A.~Butler, T.~Stuart, E.~Fisher, S.~Ghazanfar, J.~Rood, L.~Gaffney, G.~Eraslan, T.~Biancalani, E.~D. Vaishnav, R.~Conroy, D.~Procaccini, A.~Roy, A.~Pillai, M.~Brown, Z.~Galis, P.~Srinivas, A.~Pawlyk, S.~Sechi, E.~Wilder, J.~Anderson, {HuBMAP Consortium}, {Writing Group}, {Caltech-UW TMC}, {Stanford-WashU TMC}, {UCSD TMC}, {University of Florida TMC}, {Vanderbilt University TMC},
  {California Institute of Technology TTD}, {Harvard TTD}, {Purdue TTD}, {Stanford TTD}, T.~C.~V. HuBMAP~Integration, {and} Engagement (HIVE) Collaboratory: Carnegie~Mellon, T.~C. Harvard Medical~School, M.~C. Indiana University~Bloomington, I.~a. E.~C. {Pittsburgh Supercomputing Center {and} University of Pittsburgh}, C.~C. {University of South Dakota}, M.~C. New York Genome~Center, and {NIH HuBMAP Working Group}.
\newblock The human body at cellular resolution: The {{NIH Human Biomolecular Atlas Program}}.
\newblock {\em Nature}, 574(7777):187--192, Oct. 2019. doi: {{%
10\hspace{.1pt}\discretionary{.}{%
}{.}\hspace{.4pt}1038\discretionary{/}{%
}{/}s41586\discretionary{%
}{-}{-}019\discretionary{%
}{-}{-}1629\discretionary{%
}{-}{-}x}}


\bibitem{vaithilingam_dynavis_2024}
P.~Vaithilingam, E.~L. Glassman, J.~P. Inala, and C.~Wang.
\newblock {{DynaVis}}: {{Dynamically Synthesized UI Widgets}} for {{Visualization Editing}}.
\newblock In {\em Proceedings of the 2024 {{CHI Conference}} on {{Human Factors}} in {{Computing Systems}}}, {{CHI}} '24, pp. 1--17. Association for Computing Machinery, New York, NY, USA, May 2024. doi: {{%
10\hspace{.1pt}\discretionary{.}{%
}{.}\hspace{.4pt}1145\discretionary{/}{%
}{/}3613904\hspace{.1pt}\discretionary{.}{%
}{.}\hspace{.4pt}3642639}}


\bibitem{vanwijk_value_2005}
J.~Van~Wijk.
\newblock The {{Value}} of {{Visualization}}.
\newblock In {\em {{VIS}} 05. {{IEEE Visualization}}, 2005.}, pp. 79--86. IEEE, Minneapolis, MN, USA, 2005. doi: {{%
10\hspace{.1pt}\discretionary{.}{%
}{.}\hspace{.4pt}1109\discretionary{/}{%
}{/}VISUAL\hspace{.1pt}\discretionary{.}{%
}{.}\hspace{.4pt}2005\hspace{.1pt}\discretionary{.}{%
}{.}\hspace{.4pt}1532781}}


\bibitem{wang_data_2025}
C.~Wang, B.~Lee, S.~Drucker, D.~Marshall, and J.~Gao.
\newblock Data {{Formulator}} 2: {{Iterative Creation}} of {{Data Visualizations}}, with {{AI Transforming Data Along}} the {{Way}}, Feb. 2025. doi: {{%
10\hspace{.1pt}\discretionary{.}{%
}{.}\hspace{.4pt}48550\discretionary{/}{%
}{/}arXiv\hspace{.1pt}\discretionary{.}{%
}{.}\hspace{.4pt}2408\hspace{.1pt}\discretionary{.}{%
}{.}\hspace{.4pt}16119}}


\bibitem{wang_data_2024}
C.~Wang, J.~Thompson, and B.~Lee.
\newblock Data {{Formulator}}: {{AI-Powered Concept-Driven Visualization Authoring}}.
\newblock {\em IEEE Transactions on Visualization and Computer Graphics}, 30(1):1128--1138, Jan. 2024. doi: {{%
10\hspace{.1pt}\discretionary{.}{%
}{.}\hspace{.4pt}1109\discretionary{/}{%
}{/}TVCG\hspace{.1pt}\discretionary{.}{%
}{.}\hspace{.4pt}2023\hspace{.1pt}\discretionary{.}{%
}{.}\hspace{.4pt}3326585}}


\bibitem{wang_dracogpt_2025}
H.~W. Wang, M.~Gordon, L.~Battle, and J.~Heer.
\newblock {{DracoGPT}}: {{Extracting Visualization Design Preferences}} from {{Large Language Models}}.
\newblock {\em IEEE Transactions on Visualization and Computer Graphics}, 31(1):710--720, Jan. 2025. doi: {{%
10\hspace{.1pt}\discretionary{.}{%
}{.}\hspace{.4pt}1109\discretionary{/}{%
}{/}TVCG\hspace{.1pt}\discretionary{.}{%
}{.}\hspace{.4pt}2024\hspace{.1pt}\discretionary{.}{%
}{.}\hspace{.4pt}3456350}}


\bibitem{wang_natural_2022}
Y.~Wang, Z.~Hou, L.~Shen, T.~Wu, J.~Wang, H.~Huang, H.~Zhang, and D.~Zhang.
\newblock Towards {{Natural Language-Based Visualization Authoring}}.
\newblock {\em IEEE Transactions on Visualization and Computer Graphics}, pp. 1--11, 2022. doi: {{%
10\hspace{.1pt}\discretionary{.}{%
}{.}\hspace{.4pt}1109\discretionary{/}{%
}{/}TVCG\hspace{.1pt}\discretionary{.}{%
}{.}\hspace{.4pt}2022\hspace{.1pt}\discretionary{.}{%
}{.}\hspace{.4pt}3209357}}


\bibitem{wen_exploring_2025}
Z.~Wen, L.~Weng, Y.~Tang, R.~Zhang, Y.~Liu, B.~Pan, M.~Zhu, and W.~Chen.
\newblock Exploring {{Multimodal Prompt}} for {{Visualization Authoring}} with {{Large Language Models}}, Apr. 2025. doi: {{%
10\hspace{.1pt}\discretionary{.}{%
}{.}\hspace{.4pt}48550\discretionary{/}{%
}{/}arXiv\hspace{.1pt}\discretionary{.}{%
}{.}\hspace{.4pt}2504\hspace{.1pt}\discretionary{.}{%
}{.}\hspace{.4pt}13700}}


\bibitem{wongsuphasawat_voyager_2016}
K.~Wongsuphasawat, D.~Moritz, A.~Anand, J.~Mackinlay, B.~Howe, and J.~Heer.
\newblock Voyager: {{Exploratory Analysis}} via {{Faceted Browsing}} of {{Visualization Recommendations}}.
\newblock {\em IEEE Transactions on Visualization and Computer Graphics}, 22(1):649--658, Jan. 2016. doi: {{%
10\hspace{.1pt}\discretionary{.}{%
}{.}\hspace{.4pt}1109\discretionary{/}{%
}{/}TVCG\hspace{.1pt}\discretionary{.}{%
}{.}\hspace{.4pt}2015\hspace{.1pt}\discretionary{.}{%
}{.}\hspace{.4pt}2467191}}


\bibitem{wu_ai4vis_2022}
A.~Wu, Y.~Wang, X.~Shu, D.~Moritz, W.~Cui, H.~Zhang, D.~Zhang, and H.~Qu.
\newblock {{AI4VIS}}: {{Survey}} on {{Artificial Intelligence Approaches}} for {{Data Visualization}}.
\newblock {\em IEEE Transactions on Visualization and Computer Graphics}, 28(12):5049--5070, Dec. 2022. doi: {{%
10\hspace{.1pt}\discretionary{.}{%
}{.}\hspace{.4pt}1109\discretionary{/}{%
}{/}TVCG\hspace{.1pt}\discretionary{.}{%
}{.}\hspace{.4pt}2021\hspace{.1pt}\discretionary{.}{%
}{.}\hspace{.4pt}3099002}}


\bibitem{yan_knownet_2025}
Y.~Yan, Y.~Hou, Y.~Xiao, R.~Zhang, and Q.~Wang.
\newblock {{KNOWNET}}: {{Guided Health Information Seeking}} from {{LLMs}} via {{Knowledge Graph Integration}}.
\newblock {\em IEEE Transactions on Visualization and Computer Graphics}, 31(1):547--557, Jan. 2025. doi: {{%
10\hspace{.1pt}\discretionary{.}{%
}{.}\hspace{.4pt}1109\discretionary{/}{%
}{/}TVCG\hspace{.1pt}\discretionary{.}{%
}{.}\hspace{.4pt}2024\hspace{.1pt}\discretionary{.}{%
}{.}\hspace{.4pt}3456364}}


\bibitem{yu_flowsense_2020}
B.~Yu and C.~T. Silva.
\newblock {{FlowSense}}: {{A Natural Language Interface}} for {{Visual Data Exploration}} within a {{Dataflow System}}.
\newblock {\em IEEE Transactions on Visualization and Computer Graphics}, 26(1):1--11, Jan. 2020. doi: {{%
10\hspace{.1pt}\discretionary{.}{%
}{.}\hspace{.4pt}1109\discretionary{/}{%
}{/}TVCG\hspace{.1pt}\discretionary{.}{%
}{.}\hspace{.4pt}2019\hspace{.1pt}\discretionary{.}{%
}{.}\hspace{.4pt}2934668}}


\bibitem{zhu-tian_sporthesia_2023}
C.~{Zhu-Tian}, Q.~Yang, X.~Xie, J.~Beyer, H.~Xia, Y.~Wu, and H.~Pfister.
\newblock Sporthesia: {{Augmenting Sports Videos Using Natural Language}}.
\newblock {\em IEEE Transactions on Visualization and Computer Graphics}, 29(1):918--928, Jan. 2023. doi: {{%
10\hspace{.1pt}\discretionary{.}{%
}{.}\hspace{.4pt}1109\discretionary{/}{%
}{/}TVCG\hspace{.1pt}\discretionary{.}{%
}{.}\hspace{.4pt}2022\hspace{.1pt}\discretionary{.}{%
}{.}\hspace{.4pt}3209497}}


\end{thebibliography}
\end{document}